# Phonon dynamics of Zn(Mg,Cd)O alloy nanostructures and their phase segregation

Manoranjan Ghosh[1,3,a)], Nita Dilawar[2], A. K. Bandyopadhyay[2] and A.K. Raychaudhuri[1,b)]

[1]*DST Unit for Nanoscience, S.N.Bose National Centre for Basic Sciences, Block-JD, Sector-3, Salt Lake, Kolkata-700 098, INDIA*

[2]*Pressure Standards and Pressure Physics, National Physical Laboratory, Dr. K. S. Krishnan Marg, New Delhi-110012, INDIA*

[3]*Presently working at Department of Metallurgical Engineering and Materials Science, IIT Bombay, Powai, Mumbai – 400072, INDIA*

ABSTRACT: In this paper we report phonon dynamics in chemically synthesized $Zn_{1-x}Mg_xO$ ($0 \leq x \leq 0.07$) and $Zn_{1-y}Cd_yO$ ($0 \leq y \leq 0.03$) alloy nanostructures of sizes ~10 nm using non-resonant Raman and Fourier Transformed Infrared Spectroscopy (FTIR). Substitution by Mg makes the unit cell compact while Cd substitution leads to unit cell expansion. On alloying, both $A_1$(LO) and $E_1$(LO) mode of wurtzite ZnO show blue shift for $Zn_{1-x}Mg_xO$ and red shift for $Zn_{1-y}Cd_yO$ alloy nanostructures due to mass defect and volume change induced by the impurity atoms. Significant shift has been observed in $E_1$(LO) mode for $Zn_{1-x}Mg_xO$ (73 cm$^{-1}$ for $x$ = 0.07) and $Zn_{1-y}Cd_yO$ (17 cm$^{-1}$ for y = 0.03) nanostructures. The variation in Zn(Mg,Cd)-O bond length determined from the blue (red) shift of IR bands on alloying with Mg (Cd) is consistent with their respective ionic sizes and the structural changes predicted by X-ray diffraction study. However, on progressive alloying one can detect phase segregation (due to presence of interstitial Mg and Cd ions) in the alloy nanostructures for relatively higher Mg and Cd concentrations. This is confirmed by the gradual absence of the characteristic IR and Raman bands of wurtzite ZnO near 400-600 cm$^{-1}$ as well as by X-Ray and TEM studies.

[a] Email: mghosh@iitb.ac.in

[b] Email: arup@bose.res.in

## I. Introduction

In past few years, ZnO has been a widely studied subject because of its intriguing optical properties and promising application in optoelectronics.[1,2] It is a wide band gap (3.3 eV) semiconductor having wurtzite structure with lattice constants of a = b = 3.25 Å and c = 5.207 Å.[3,4] The fundamental absorption even at room temperature shows excitonic contribution due to its large excitonic binding energy (~ 60meV).[5] Majority of the previous reports on ZnO and ZnO based nanostructures are focused on their optical properties such as transmission and photoluminescence.[1-7] Like other semiconductors, ZnO nanostructures show possibility of tuning its properties by alloying and exhibit flexible band gap and emission energies by substituting (alloying) bivalent metals like Cd and Mg in place of Zn. While Cd is known to reduce the band gap[7], Mg substitution leads to enhancement of band gap[6]. This raises the possibility of band gap engineered heterostructures for optoelectronic applications in the UV range which can serve as alternative for the Gr-III nitride based optoelectronic material.

On the other hand, limited studies are available on understanding the specific phonon spectrum of ZnO based nanostructures which is important in developing optoelectronic devices.[8,9] Specially, phonon dynamics of band gap engineered alloyed ZnO nanostructures such as $Zn_{1-x}Mg_xO$ and $Zn_{1-x}Cd_xO$ is a rarely studied area and requires further study.[10,11,12] Lattice vibrational properties of ZnMgO films grown by vapour phase methods have been investigated.[13,14] After successful fabrication of colloidal Zn(Mg,Cd)O alloy nanostructures (by chemical synthesis method) with tunable transmission and photoluminescence properties,[6,7] it is possible to study the phonon dynamics of polycrystalline ZnO and Zn(Mg,Cd)O alloy in the size regime of ~ 10 nm.

In this work we describe non-resonant Raman and FTIR investigation of optical phonons in ZnO and Zn(Mg,Cd)O alloy nanostructures (size ~ 10 nm). The Raman and IR active vibrational modes are identified and the shift in $E_1(LO)$ mode of $Zn_{1-x}Mg_xO$ and $Zn_{1-y}Cd_yO$ alloys is investigated. It appears that for higher precursor concentrations, the alloying leads to phase segregation predicted by additional peaks in the XRD pattern and absence of characteristic IR bands.

**II. Experimental**

Crystalline $Zn_{1-x}Mg_xO$ and $Zn_{1-y}Cd_yO$ alloy nanostructures have been synthesized via chemical route under ambient and 54 atm. pressure respectively. The detail of the method of synthesis is described elsewhere.[6,7] In Table I we show the size, morphology and chemical compositions of all the samples investigated in this work. The sizes are determined by transmission electron microscope (TEM) imaging as well as Williamson Hall analysis of the XRD results.[15] The root mean square (rms) distribution in size ($d$) of the spherical particles can be quantified as $\Delta d_{rms}/d \approx 10\%$ for all the samples. The commercial ZnO powder (referred as bulk ZnO and is used as reference) which show mixed morphologies like square pillars, rods and cubes of size 300 nm and particles of size ~ 100 nm attached with them.

The Cd and Mg content $x$ was determined by Inductive Coupled Plasma-Atomic Emission Spectroscopy (ICP-AES) analysis which is an accurate method with calibration done using the available standards. Nanostructures and their lattice fringes were imaged by a JEOL High Resolution TEM (HRTEM) working at 200 KeV and the XRD data were obtained using a Philips X' pert-Pro X-ray diffractometer. The room temperature non-resonant Raman measurement has been carried out by a home made Raman spectrometer consisting of a Jobin Yvon-Spex (HR640) monochromator with a notch filter (514.5 nm) and $Ar^+$ ion laser.[16] Fourier Transform Infrared (FTIR) spectra of the powdered samples in KBr pellet have been collected by a JASCO FT/IR-6300 spectrometer. Both Raman and IR spectra were taken at room temperature.

## III. Results and discussion

### A. TEM and XRD study

The representative TEM images of the undoped ZnO (x=0) nanoparticles, commercial ZnO powder, $Zn_{1-x}Mg_xO$ (x= 0.02) and $Zn_{1-y}Cd_yO$ (y = 0.02) nanostructures have been shown in Fig. 1 (a), (b) (c) and (d) respectively. The size of the nanostructures seen by the TEM images agrees well with that determined from the XRD results and lies within the weak confinement regime (10 – 15 nm). The alloy nanostructures prepared by this chemical route retains the wurtzite structure of parent ZnO below a certain precursor concentrations of Mg and Cd as seen by the XRD and Fast Fourier Transform (FFT) of the HRTEM images.[6,7] However, the average lattice constants (c-axis length) increases and decreases with dopant concentrations for $Zn_{1-x}Mg_xO$ and $Zn_{1-y}Cd_yO$ respectively, due to the mismatch in ionic radii of Zn (0.74 Å), Mg (0.65 Å) and Cd (0.97 Å) as seen by the Rietveld analysis of the XRD data described in detail elsewhere.[6,7] Further increase in precursor concentration (for Mg and Cd alloying) leads to phase segregation as shown in Fig. 2. Distinct peaks for $Mg(OH)_2$ and CdO appears after the addition of 40 wt.% and 20 wt.% of $Mg(CH_3COO)_2.4H_2O$ and $Cd(CH_3COO)_2.2H_2O$ respectively into the $Zn(CH_3COO)_2.2H_2O$ solution during the synthesis process. It should be noted that phase segregation appears very quickly in case of $Zn_{1-y}Cd_yO$ alloy due to the existence of higher lattice mismatch ($\Delta R_{Cd}$ = 0.23 Å) compared to the $Zn_{1-x}Mg_xO$ alloy ($\Delta R_{Mg}$ = 0.09 Å).

**B. Raman study**

Group theory predicts that near the centre of the Brillouin zone, there is an $A_1$ branch, a doubly degenerate $E_1$ branch, two doubly degenerate $E_2$ branches, and two B branches of ZnO with $C_{6V}$ symmetry. The A1 and E1 branches are both Raman and infrared active; the E2 branches are Raman-active only, while the B branches are inactive.[17] The room temperature non-resonant Raman spectra of polycrystalline ZnO having orientation in all directions exhibit all the characteristic Raman peaks as shown in Fig 3 (a) for commercially available ZnO powder of particle size > 100 nm. In Table II, we show that the frequencies of the characteristic Raman active modes for commercially available polycrystalline ZnO matches well with the earlier results. Due to the tilted geometry of our samples, prominent occurrence of all the peaks for all the samples is not seen. As for example, $E_1$ (TO) modes strongly appears for $Zn_{1-x}Mg_xO$ (with x = 0.02) and $Zn_{1-y}Cd_yO$ samples. Again the $A_1$ (TO) mode appears clearly in case of bulk ZnO powder and lower Mg content sample. In addition, a broad peak at 335 cm$^{-1}$ is observed which is attributed to the second order Raman processes.

**Effect of alloying:** As indicated in the above discussion, the wurtzite structure has two symmetry types of Raman active LO modes viz $A_1$ (LO) and $E_1$ (LO) which arises depending on the experimental geometry. In crystallites of tilted orientation (as in powders), the two LO modes are expected to interact and create a single mode of a mixed $A_1$−$E_1$ symmetry known as the quasi-LO mode.[10] Previous studies concerning the phonon modes of wurtzite $Zn_{1-x}Mg_xO$ thin films and powders showed that the $E_1(LO)$ frequency of the film exhibited a significant blue shift while the $A_1(LO)$ frequency was unaffected by the increasing Mg concentration.[11,18] In this work, the mixed $A_1$–$E_1$ mode has been resolved into two separates modes viz. $A_1$ (LO) and $E_1$ (LO) by Gaussian fitting as demonstrated in Fig. 3 (c) for $Zn_{1-x}Mg_xO$ sample with x = 0.02. We have observed a shift in both $A_1$ (LO) and $E_1$ (LO) modes due to alloying of ZnO by Mg and Cd [Fig. 3(b)]. Also the broad peak appearing in the low frequency region due to multi-phonon processes becomes broader for alloy nanostructures. Interestingly, the frequency of $E_2$ (high) mode remains unaffected by the alloying which was not clearly mentioned in the previous reports.[11,18] Another very important issue in alloying of ZnO is the phase segregation. As can be seen from the XRD data (Fig. 2), if more than 40 wt.% of magnesium acetate and 20 wt.% of cadmium acetate are added to the zinc acetate, phase segregation occurs with $Mg(OH)_2$ and CdO coexisting with $Zn_{1-x}Mg_xO$ and $Zn_{1-y}Cd_yO$ phases respectively. Raman spectra also shows the effect of phase segregation in the case of $Zn_{1-x}Mg_xO$ (x >0.17) sample as indicated in Fig. 3(b). Characteristic Raman modes are averaged out due to the lack of long range order and the Raman signal partly bears the signature of an amorphous material. Although the $Zn_{1-y}Cd_yO$ alloy nanostructures do not show phase segregation up to y = 0.03 (determined by XRD results), the broad Raman spectrum in the low frequency region indicates disorder in the sample.

The lattices of semiconductors can be perturbed by the addition of foreign atoms or by the application of stress. These perturbations change the force constant and mass matrices of the lattice vibrations. Hence a detailed knowledge of the lattice dynamics of semiconductors is essential in treating their common perturbations. In case of isoelectronic substitution of Mg and Cd into ZnO lattice, one can assume that the mass defect (M) and the volume change resulting from the difference in atomic radii (R) of the impurity and the host atoms is uniformly distributed over all the atoms of the crystal. The only predicted effect is a frequency shift induced by the perturbations such as mass defect and volume change.[19] These perturbations eventually depend on the concentration of the impurities. Also we have come across that the mass defect ($\Delta M$) and difference in radii ($\Delta R$) are -41.085 and -0.09 ˚A respectively for Mg substitution and that of 47.02 and 0.23 ˚A for Cd substitution. The negative (+ve) sign applied in the difference values for Mg (Cd) substitution implies that the Mg (Cd) atom is lighter (heavier) and smaller (bigger) in diameter than the Zn atom and leads to compaction (expansion) of lattice. Therefore a particular Raman mode shifts to higher (lower) frequency side as a result of Mg (Cd) incorporation into Zn lattice site. Because of the alloying, the $E_1$ (LO) mode shows blue shift up to 73 $cm^{-1}$ (for x = 0.07) and red shift up to 17 $cm^{-1}$ (for y = 0.03) for $Zn_{1-x}Mg_xO$ and $Zn_{1-y}Cd_yO$ alloy nanostructures respectively [Fig. 3 (d) and Table III]. The broadening of the multi-phonon Raman modes at higher Cd and Mg concentrations can be associated with the activation/formation of a number of structural defects in doped ZnO.[20] As we have seen, these chemically synthesized alloy nanostructures show phase segregation after a certain concentration of the impurities. Generally, the other alternative alloying process results in the nearly random spatial distribution of the intermixing atomic species. This randomness removes the translational symmetry of the crystal, so that Raman spectra from semiconductor alloys might be expected to be similar to the Raman spectra from amorphous semiconductors, which are roughly proportional to the density of phonon states. This is reflected in the broad peak observed in the low frequency region (300-400 $cm^{-1}$) for the alloy nanostructures. The broadening is more prominent in case of $Zn_{1-y}Cd_yO$ alloy nanosructures because the atomic radii mismatch is higher in case of Zn and Cd (0.23 Å) compared to that of Zn and Mg (0.09 Å) as mentioned earlier in this section. So, the substitution of Cd into ZnO lattice is difficult and intermixing of two atomic species is highly probable. However, alloying is in general a much gentler perturbation than amorphization, as seen from the X-ray diffraction studies of our samples which clearly indicates the existence of a remaining order and a well defined average lattice constant.

**C. Fourier Transform Infrared Spectroscopy (FTIR) study**

Infrared transmission spectra of ZnO and its alloy nanoparticles embedded in KBr powder media have been measured in the wavelength range 375 to 4000 cm$^{-1}$ (which is much higher than the particle diameter ~10 nm) to provide an additional support to the observation made by Raman spectroscopy. The IR active optical phonon modes of ZnO show a characteristic broad reststrahlen band in the spectral range of 300-600 cm$^{-1}$.[21] Higher wave number peaks correspond to vibration modes of various impurities such as hydroxyl, carboxylate and alkane in the materials.[22] In this work, within the displayed frequency range, phase segregated ZnMgO and ZnCdO alloy nanostructures show two impurity bands near 600-1000 cm$^{-1}$ due to C-OH bond. The impurity peaks are present in almost all the previous reports of ZnO nanostructures grown by chemical as well as physical deposition method.[22,23] The sources of these impurities are the surface adsorbed organic precursors employed during the synthesis process. The effect of these impurities is prominent for smaller nanoparticles due to high surface to volume ratio and the impurity peaks gradually vanish for bigger particles.

As mentioned earlier in this paper, only $A_1$ and $E_1$ modes of ZnO are IR active. Investigation done by the polarized light on $Zn_{0.8}Mg_{0.2}O$ thin film gives rise to four distinct vibrational modes, such as $A_1(TO)$, $E_1(TO)$, $A_1(LO)$ and $E_1(LO)$ at 384 cm$^{-1}$, 417 cm$^{-1}$, 505 cm$^{-1}$ and 586 cm$^{-1}$.[24] Characteristic vibration modes are not distinct for nanostructures of size ~ 10 nm adopted in this work due to absorption by the surface states at lower sizes. For clarity, the spectrum for bulk ZnO (~size > 100 nm) is shown where IR bands in the range of 375 cm$^{-1}$- 600 cm$^{-1}$ have been indicated by P1, P2, P3 and P4 (four Gaussians fitting shown in Fig. 4). In agreement with the previous studies,[21,24] the P1 (~390 cm$^{-1}$) and P2 (~440 cm$^{-1}$) bands can be assigned as $A_1(TO)$ and $E_1(TO)$ respectively [Table IV]. It has been reported that the interaction between electromagnetic (EM) radiation and the particles depends on the size, shape, and state of aggregation of the crystal.[21] Since the particle size in this study is much lower than the EM radiation, surface phonon modes (SPM) can be observed in the IR spectra as well. For spherical particles, the value of the first order (n=1) SPM is given by,[25]

$$\omega_o^2 = \frac{\varepsilon_0 + 2\varepsilon_m}{\varepsilon_\infty + 2\varepsilon_m}\omega_{TO}^2$$

Here, $\varepsilon_0$ = 8.4, $\varepsilon_\alpha$ = 3.72 (high frequency dielectric constant of ZnO), $\varepsilon_m$ = 2.36 (dielectric constant of KBr medium). So, the bands around 495 cm$^{-1}$ (P3) and 540 cm$^{-1}$ (P4) can be identified as the SPM absorption corresponding to the TO bands around 395 cm$^{-1}$ and 435 cm$^{-1}$ respectively. The IR bands discussed above show a systemic variation in the vibration frequencies after progressive alloying. As shown in Fig. 5, all the IR modes shift to the blue and red side of the spectrum for $Zn_{1-x}Mg_xO$ and $Zn_{1-y}Cd_yO$ alloys respectively.

The shift in band positions on doping of Mg and Cd can be correlated to their respective ionic sizes as well as the structural changes induced due to doping. The vibrational frequency of Zn (M) - O bonds of $Zn_{1-n}M_nO$ (M = Mg/Cd, n = x for Mg and n = y for Cd) can be determined from band position of $E_1(TO)$ mode by the equation,[26,27]

$$\bar{v} = \frac{1}{2\pi c}\left[\frac{k}{\mu}\right]^{1/2} \quad \ldots\ldots\ldots (1)$$

where, $\bar{v}$ is the wave number, c the velocity of light, k average force constant of the Zn (M) - O bond and µ the effective mass of the bond which is given by the following:

$$\mu = \frac{M_O \times [xM_M + (1-x)M_{Zn}]}{M_O + [xM_M + (1-x)M_{Zn}]} \quad \ldots\ldots (2)$$

where, $M_o$, $M_{Zn}$ and $M_M$ are the atomic weights of O, Zn and M (Mg/Cd) respectively. Further, force constant (k) can be related to the average Zn (M) - O bond length (r in Å) by the equation,[26]

$$k = \frac{17}{r^3} \quad \ldots\ldots\ldots (3)$$

Thus, the approximate values of effective mass, the force constant and bond length were calculated by Eqs. (1) - (3), and shown in the Table V. It is seen from Table V that the effective mass of Zn (Mg)-O bond decreases (increases) with Mg (Cd) substitution because of lower (higher) atomic weight of Mg (Cd) than Zn. Also the average force constant decreases (increases) with Cd (Mg) substitution which results increase (decrease) in the average Zn (Cd)-O [Zn (Mg)-O] bond length. These changes in local structure parameters are therefore linked with the substitution of Mg and Cd into ZnO lattice and their respective ionic radii. The variation in bond length due to Mg and Cd substitution is consistent with the trend found by Rietveld analysis of X-ray and Synchrotron X-ray data from polycrystalline ZnMgO nanomaterials although a little change in cell volume is observed.[28,29] In contrast, both cell volume and Zn (Cd) - O bond length of ZnCdO nanomaterials increases significantly.[7] Therefore, the shift in IR modes due to alloying by Mg and Cd corroborates the structural studies by complementary techniques. It should be noted that similar to the Raman, FTIR spectra also carries the signature of phase segregation. Progressive alloying eventually leads to phase segregation which results in a flat FTIR spectrum in the range of 300 -600 cm$^{-1}$.

## IV. Conclusions

An analysis of the optical phonon modes in polycrystalline ZnO and Zn(Mg,Cd)O alloy nanostructures has been documented. On alloying, both $A_1$ (LO) and $E_1$ (LO) mode of wurtzite Zn(Mg,Cd)O nanostructures show blue shift for $Zn_{1-x}Mg_xO$ and red shift for $Zn_{1-y}Cd_yO$ alloy nanostructures. The shift in the phonon frequencies has been explained on the basis of induced perturbation by the mass defect and volume change due to incorporation of impurity atoms. The IR modes also shift towards higher and lower wave numbers for $Zn_{1-x}Mg_xO$ (x = 0.07) and $Zn_{1-y}Cd_yO$ (y = 0.03) alloy nanostructures respectively. The variation in bond length predicted by the position of IR bands is consistent with the structural parameters found from X-ray diffraction study. Signature of phase segregation in the alloy nanostructures is reflected by the absence of the characteristic IR and Raman bands of wurtzite ZnO in the range of 400-600 cm$^{-1}$.

**ACKNOWLEDGEMENT**

The authors acknowledge the technical help from TEM facility at the Indian association for the Cultivation of Science and the Department of Science and Technology for providing the financial support.

## References

[1] A. van Dijken, J. Makkinje, and A. Meijerink, J. of luminescence **92**, 323 (2001).

[2] Sun X. W, Yu S. F, Xu C. X, Yuen C, Chen B. J, and Li S, Jpn. J. Appl. Phys. Part 2 **42,** L1229 (2003).

[3] D. M. Bagnall, Y. F. Chen, Z. Zhu, T. Yao, S. Koyama, M. Y. Shen, and T. Goto, Appl. Phys. Lett. **70**, 2230 (1997).

[4] S. J. Pearton et. al, J. Appl. Phys. **93**, 1 (2003).

[5] Z. K. Tang, G. K. L. Wong, P. Yu, M. Kawasaki, A. Ohtomo, H. Koinuma, and Y. Segawa, Appl. Phys. Lett. **72,** 3270 (1998).

[6] M. Ghosh, and A. K. Raychaudhuri, J. Appl. Phys. **100**, 034315 (2006).

[7] M. Ghosh, and A. K. Raychaudhuri, Nanotechnology **18**, 115618 (2007).

[8] K. A. Alim, V. A. Fonoberov, M. Shamsa, and A. A. Balandin, J. Appl. Phys. **97**, 124313 (2005).

[9] H. M. Cheng, H. C. Hsu, Y. K. Tseng, L. J. Lin, and W. F. Hsieh, J. Phys. Chem. B **109**, (2005) 8749.

[10] L. Bergman, J. L. Morrison, X. B. Chen, J. Huso, and H. Hoeck, Appl. Phys. Lett. **88**, 023103 (2006).

[11] C. Bundesmann, A. Rahm, M. Lorenz, M. Grundman, and M. Schubert, J. Appl. Phys. **99**, 113504 (2006).

[12] F. Wang, H. He, Z. Ye, L. Zhu, H. Tang, and Y. Zhang, J. Phys. D: Appl. Phys**. 38**, 2919 (2005).

[13]J.F. Konga, W.Z. Shen, Y.W. Zhang, X.M. Li, and Q.X. Guo, Solid State Communications **149**, 10 (2009).

[14]A. I. Belogorokhov, A. Y. Polyakov, N. B. Smirnov, A. V. Govorkov, E. A. Kozhukhova, H. S. Kim, D. P. Norton, and S. J. Pearton, Appl. Phys. Lett. **90**, 192110 (2007).

[15]G. K. Williamson, and W. H. Hall, Acta Metall. **1**, 22 (1953).

[16]A. K. Bandyopadhyay, N. Dilawar, A. Vijayakumar, D. Varandani, and D. Singh, Bull. Mater. Sci. **21**, 433 (1998).

[17]T. C. Damen, S. P. S. Porto, B. Tell, Phys. Rev. B **142**, 570 (1966).

[18]J. Huso, L. Morrison, H. Hoeck, E. Casey, L. Bergman, T. D. Pounds, and M. G. Norton, Appl. Phys. Lett. **91**, 111906 (2007).

[19] W. H. Weber, and R. Merlin, Raman Scattering in Material Science, Springer 2000, ISBN 3540672230.

[20]X. Wang, J. Xu, B. Zhang, H. Yu, J. Wang, X. Zhang, J. Yu and Q. Li, Adv. Mater. **18**, 2476 (2006).

[21]B Cheng, Y Xiao, G Wu, and L Zhang, Appl. Phys. Lett. **84**, 2004 (416).

[22]G. Xiong, U. Pal, and J. G. Serrano, J. Appl. Phys. **101**, 024317 (2007).

[23]Y. J. Kwon, K. H. Kim, C. S. Lim, and K. B. Shim, Journal of Ceramic Processing Research **3**, 146 (2002).

[24]C. Bundesmann, M. Schubert, D. Spemann, T. Butz, M. Lorenz, E. M. Kaidashev, M. Grundmann, N. Ashkenov, H. Neumann, and G. Wagner, Appl. Phys. Lett. **81**, 2376 (2002).

[25]K. Yamamoto, C-D Tran, H. Shimizu, and K. Abe, J. Phys. Soc. Jpn. **42**, 587 (1977).

[26]R. A. EL-Mallawany, Infrared Phys. **29**, 781 (1989).

[27] K. C. Barick and D. Bahadur, J. Nanosci. Nanotechnol. **8**, 4263 (2008).

[28] Z. K. Heiba and L. Arda, Cryst. Res. Technol. **44**, 845 (2009).

[29] Y-I Kim, K. Page, and R Seshadri, Appl. Phys. Lett. **90**, 101904 (2007).

TABLE I. Size, morphology and chemical compositions of the nanostructures investigated.

| Samples | | Morphology | Size (nm) |
|---|---|---|---|
| ZnO | Bulk | Square pillar or rod | >100 |
| $Zn_{1-x}Mg_xO$ | X=0 | Spherical | 10 |
| | X= 0.02 | Spherical | 13 |
| | X= 0.07 | Spherical | 12 |
| | X=0.40 | Spherical | 14 |
| $Zn_{1-y}Cd_yO$ | Y= 0.02 | Spherical and rods | 11 |
| | Y=0.03 | Spherical and rods | 12 |

TABLE II. Assignments of characteristic Raman modes for bulk ZnO powder.

| Symmetry character | Bulk ZnO (cm$^{-1}$) (our work) | Ref. 15 (cm$^{-1}$) |
|---|---|---|
| $E_2$ (high) | 438 | 437 |
| $A_1$ (TO) | 381 | 380 |
| $E_1$ (TO) | 409 | 407 |
| $A_1$ (LO) | 571 | 574 |
| $E_1$ (LO) | 583 | 583 |
| Multi-phonon processes | 335 | 334 |

TABLE III. Raman modes of ZnO and its alloy nanostructures.

| | $E_2$ (high) (cm$^{-1}$) | $A_1$ (LO) (cm$^{-1}$) | $E_1$ (LO) (cm$^{-1}$) | Shift in $E_1$ (LO) (cm$^{-1}$) | Multi-phonon peaks (cm$^{-1}$) | $A_1$ (TO) (cm$^{-1}$) |
|---|---|---|---|---|---|---|
| ZnO bulk | 438 | 571 | 583 | | 335 | 381 |
| X=0 | 437 | 579 | 587 | | 346 | 381 |
| X = 0.02 | 437 | 593 | 618 | 31 | 355 | 388 |
| X = 0.07 | 437 | 651 | 663 | 76 | 346 | 382 |
| Y = 0.02 | 437 | 564 | 575 | 12 | | |
| Y = 0.03 | 436 | 556 | 569 | 17 | | |

TABLE IV. IR active vibration modes for ZnO and its alloy nanostructures.

| Samples | $A_1$(TO) (cm$^{-1}$) | $E_1$(TO) (cm$^{-1}$) | SPM[$A_1$(TO)] (cm$^{-1}$) | SPM[$E_1$(TO)] (cm$^{-1}$) |
|---|---|---|---|---|
| Bulk ZnO | 393 | 435 | 493 | 540 |
| X=0 | 395 | 450 | 496 | 533 |
| X= 0.02 | 399 | 452 | 499 | 534 |
| X= 0.07 | 405 | 462 | 503 | 539 |
| Y= 0.02 | 391 | 427 | 482 | 535 |
| Y=0.03 | 380 | 413 | 457 | 518 |

TABLE V. The IR band and local structure data of Zn (M) - O bonds of $Zn_{1-n}M_nO$ (M = Mg/Cd, n = x for Mg and n = y for Cd) samples.

| Samples | Wave number (cm$^{-1}$) | Effective mass (atomic weight) | Force constant (Ncm$^{-1}$) | Bond length (Å) |
| --- | --- | --- | --- | --- |
| x=0.07 | 462 | 12.739 | 1.604 | 2.196 |
| x = 0.02 | 452 | 12.823 | 1.546 | 2.224 |
| x = 0 | 450 | 12.855 | 1.536 | 2.229 |
| y = 0.02 | 427 | 12.891 | 1.387 | 2.306 |
| y = 0.03 | 413 | 12.908 | 1.299 | 2.357 |

**Figure Captions**

FIG. 1. Representative TEM images of (a) undoped ZnO nanoparticles of size~10 nm, (b) commercial ZnO of size 100 nm-500 nm (used as reference), (c) $Zn_{1-x}Mg_xO$ (x = 0.02) and (d) $Zn_{1-y}Cd_yO$ (y = 0.02) nanostructures.

FIG. 2. (Color online) XRD data of selected samples as indicated on the graph. All the samples show wurtzite ZnO peaks as indexed for bulk ZnO. Distinct peaks for $Mg(OH)_2$ and CdO appears (as indexed on the respective curve) in case of phase segregated $Zn_{1-x}Mg_xO$ and $Zn_{1-y}Cd_yO$ nanostructures.

FIG. 3. (Color online) Non-resonant Raman spectra of (a) Bulk ZnO powder (b) series of $Zn_{1-x}Mg_xO$ and $Zn_{1-y}Cd_yO$ alloy nanostructures. Assigned Raman modes are indicated on the graph. A sharp peak at 521 $cm^{-1}$ is originated from the glass substrate (c) The LO mode of $Zn_{1-x}Mg_xO$ (x = 0.02) nanostructures is fitted by two Gaussian to resolve $A_1$ (LO) and $E_1$ (LO) modes of ZnO. Observed shift in the $E_1$ (LO) mode with Mg and Cd concentration is plotted in (d).

FIG. 4. (Color online) The FTIR spectra of $Zn_{1-x}Mg_xO$, $Zn_{1-y}Cd_yO$ and phase segregated MgZnO alloy as indicated on the graph.

FIG. 5. (Color online) Variation in the IR modes of $Zn_{1-x}Mg_xO$ and $Zn_{1-y}Cd_yO$ alloy nanoparticles as indicated on the graph.

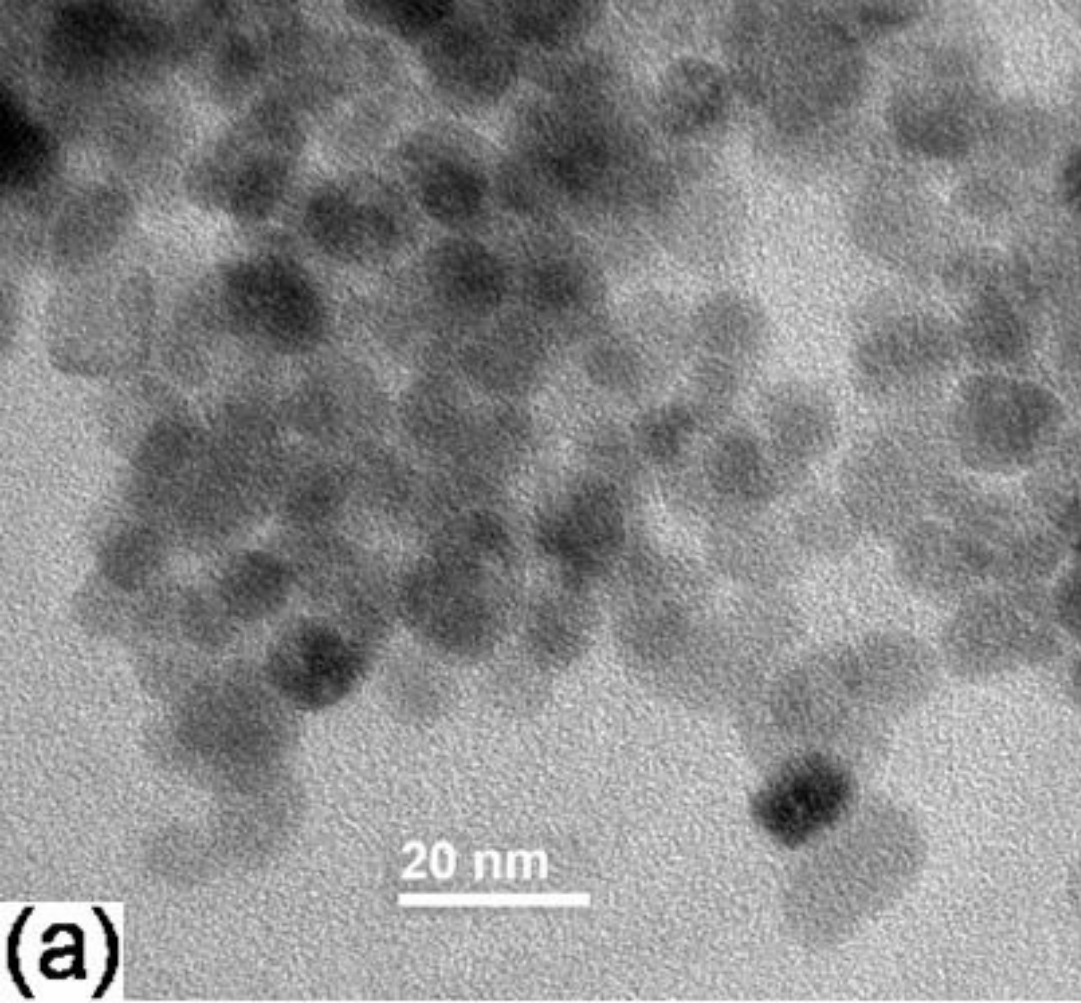
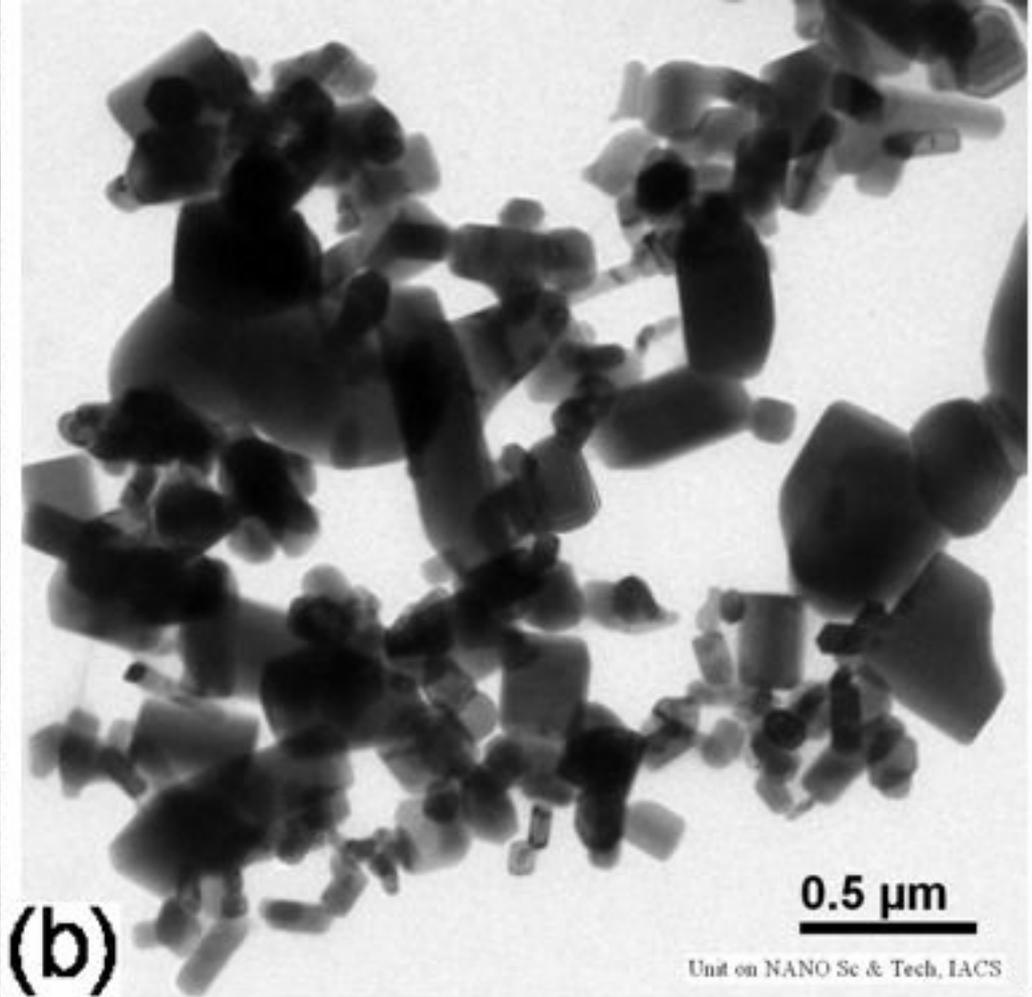
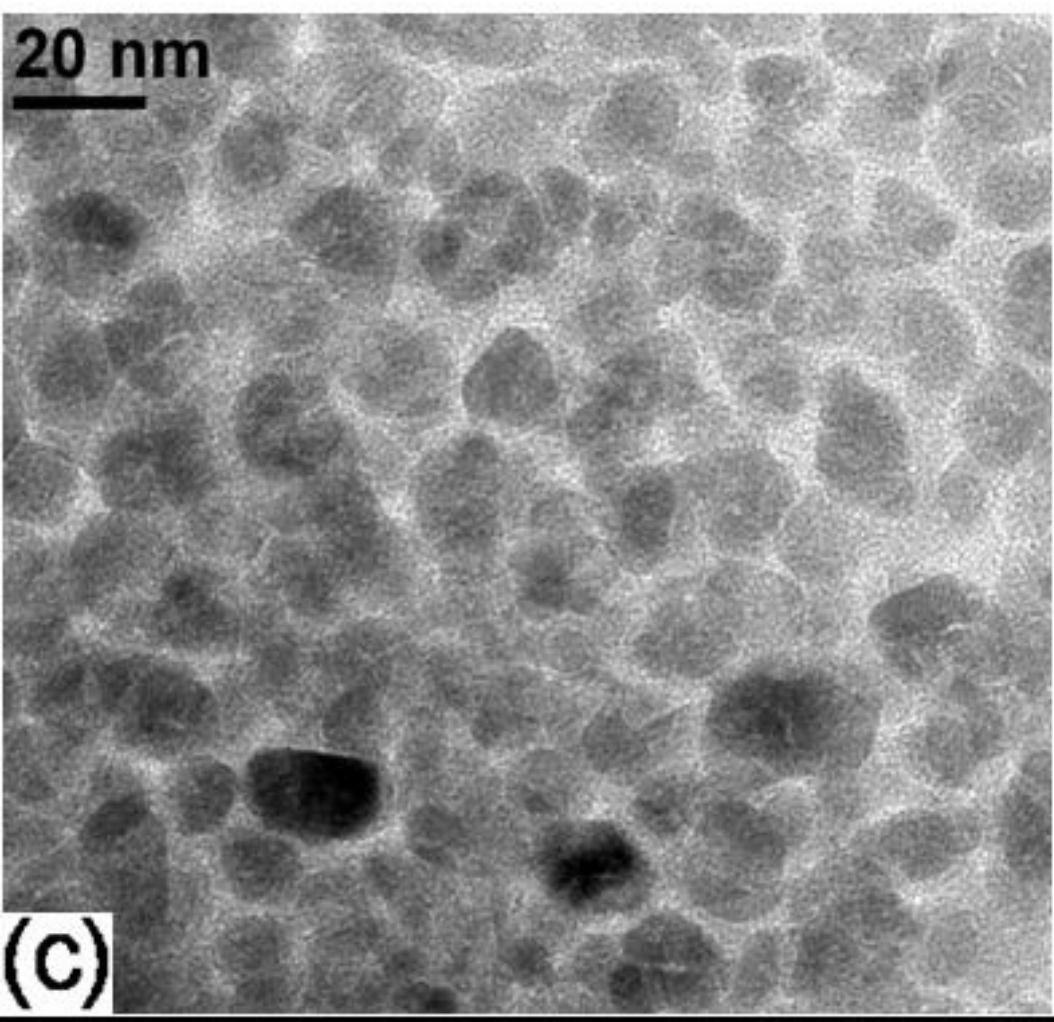
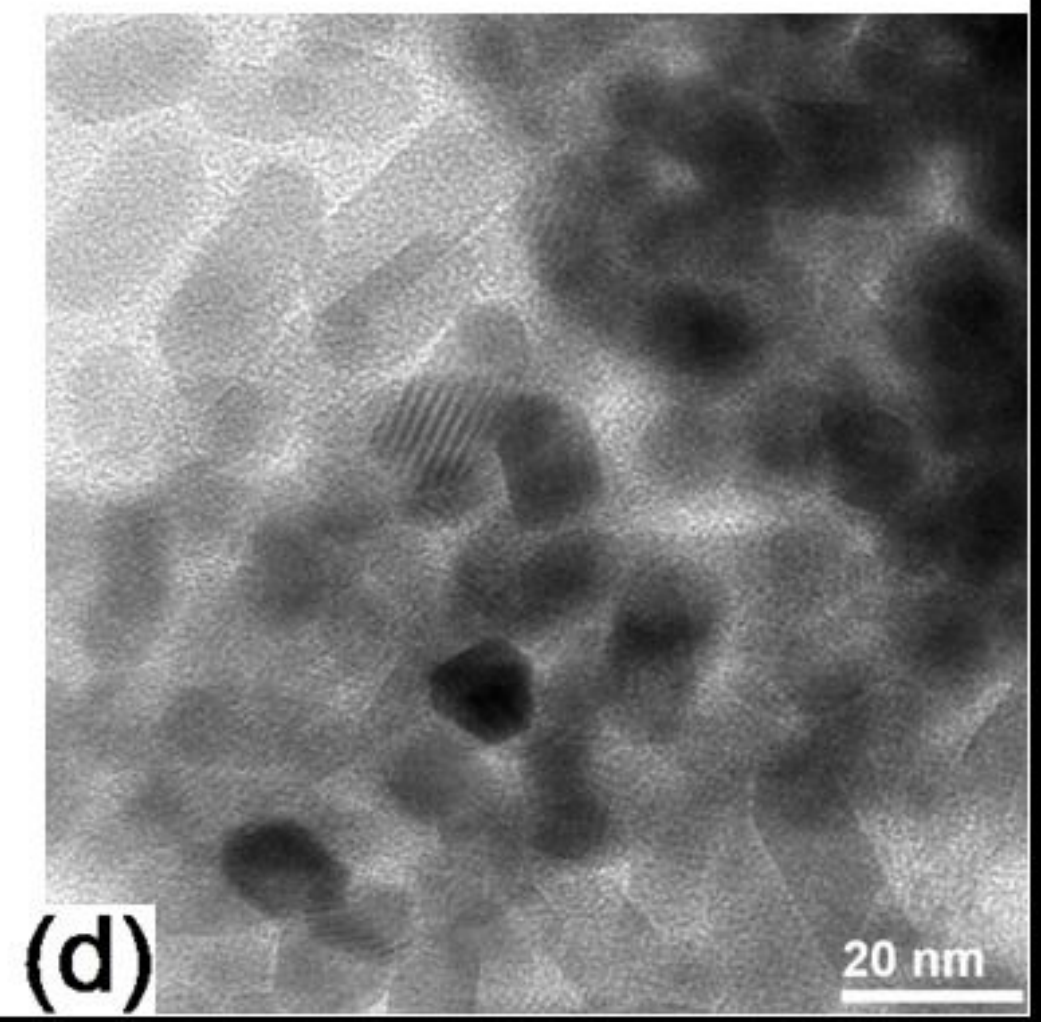

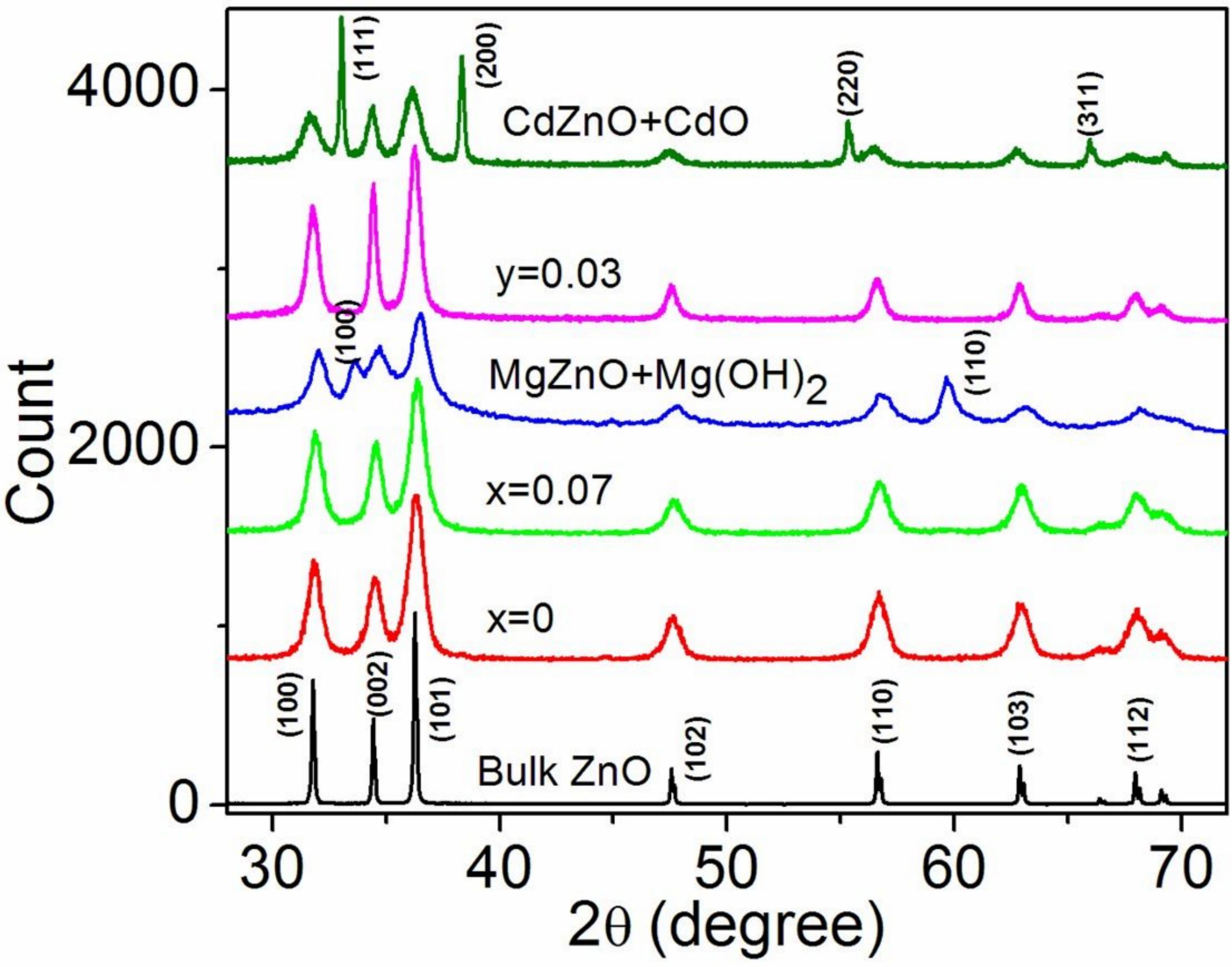

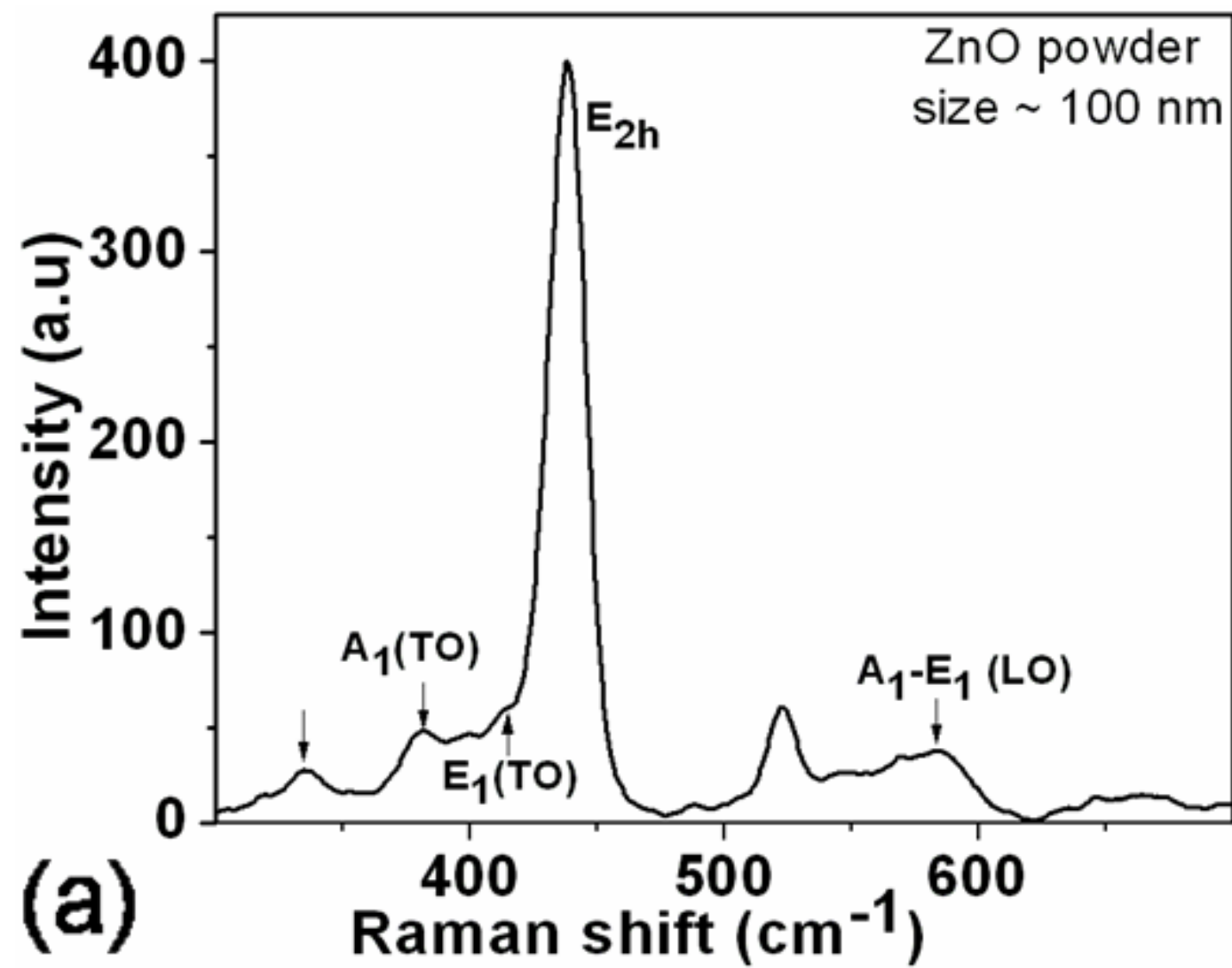
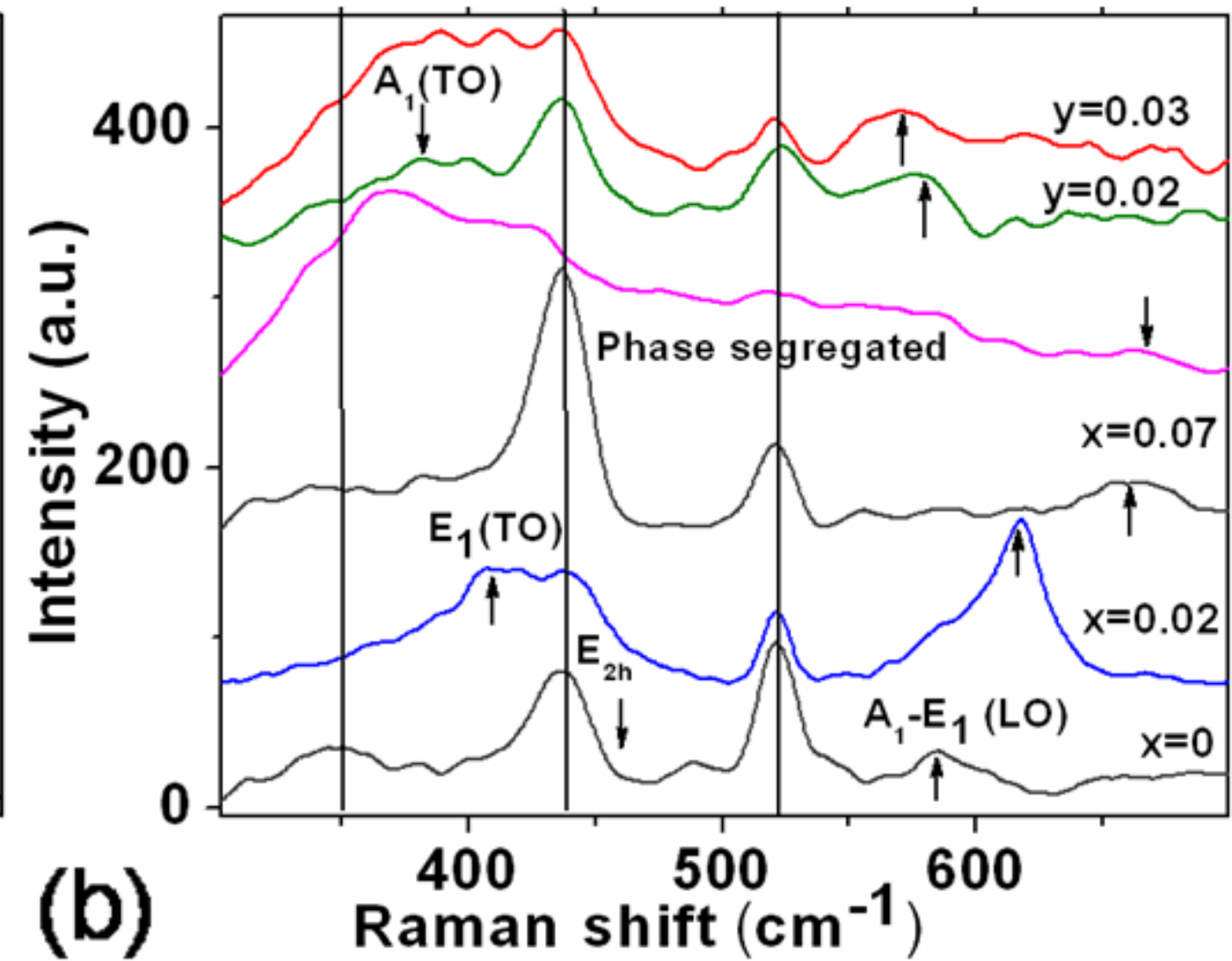
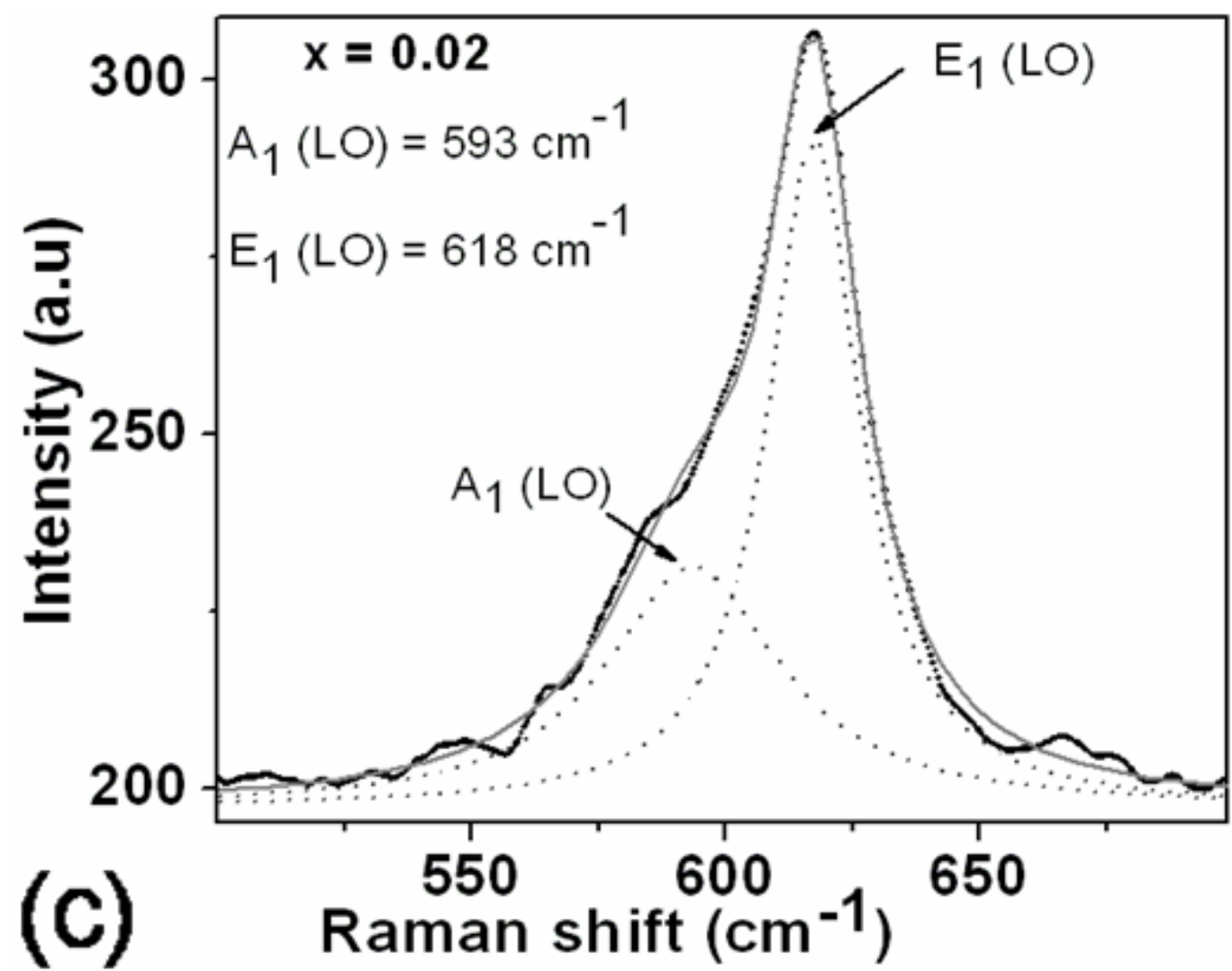
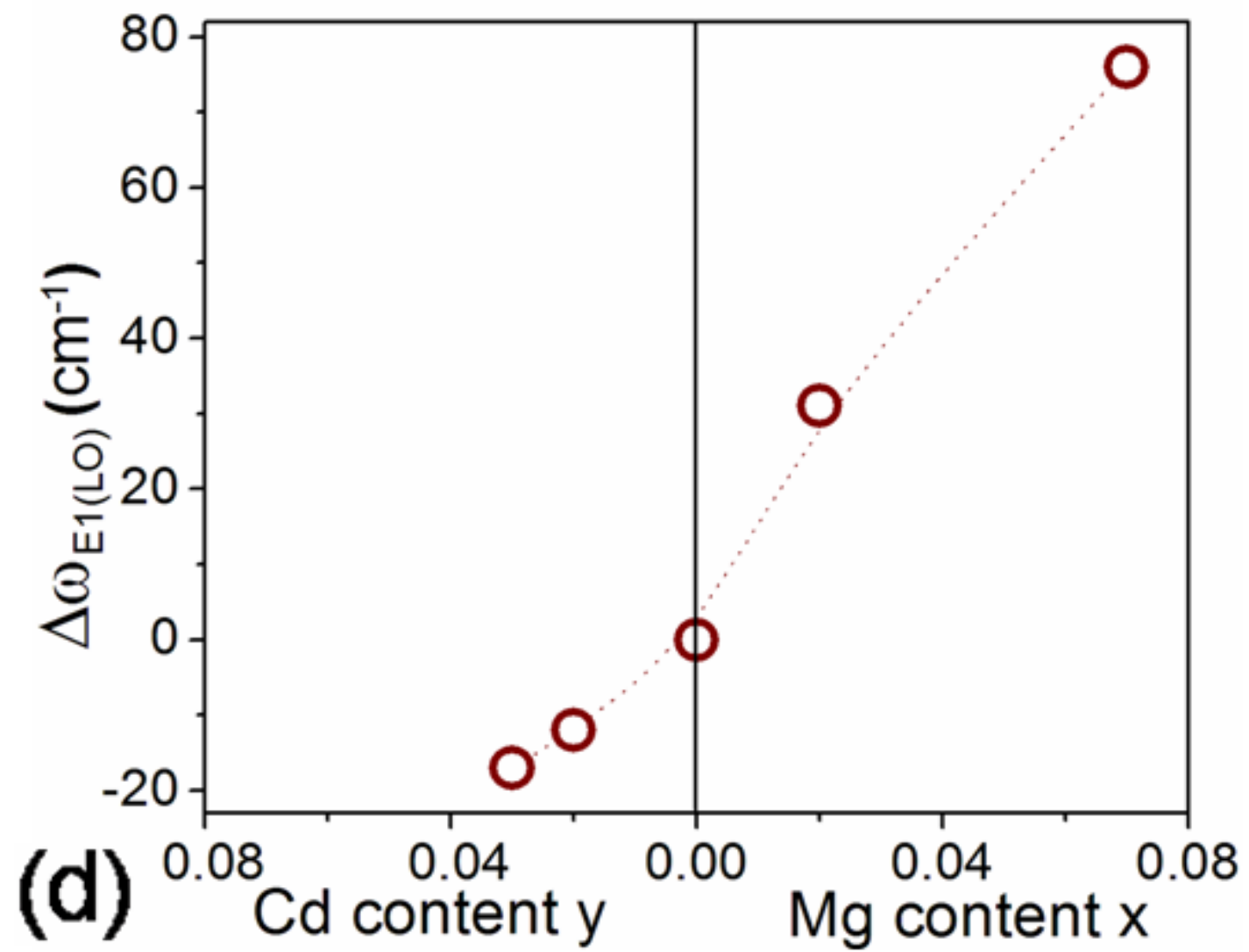

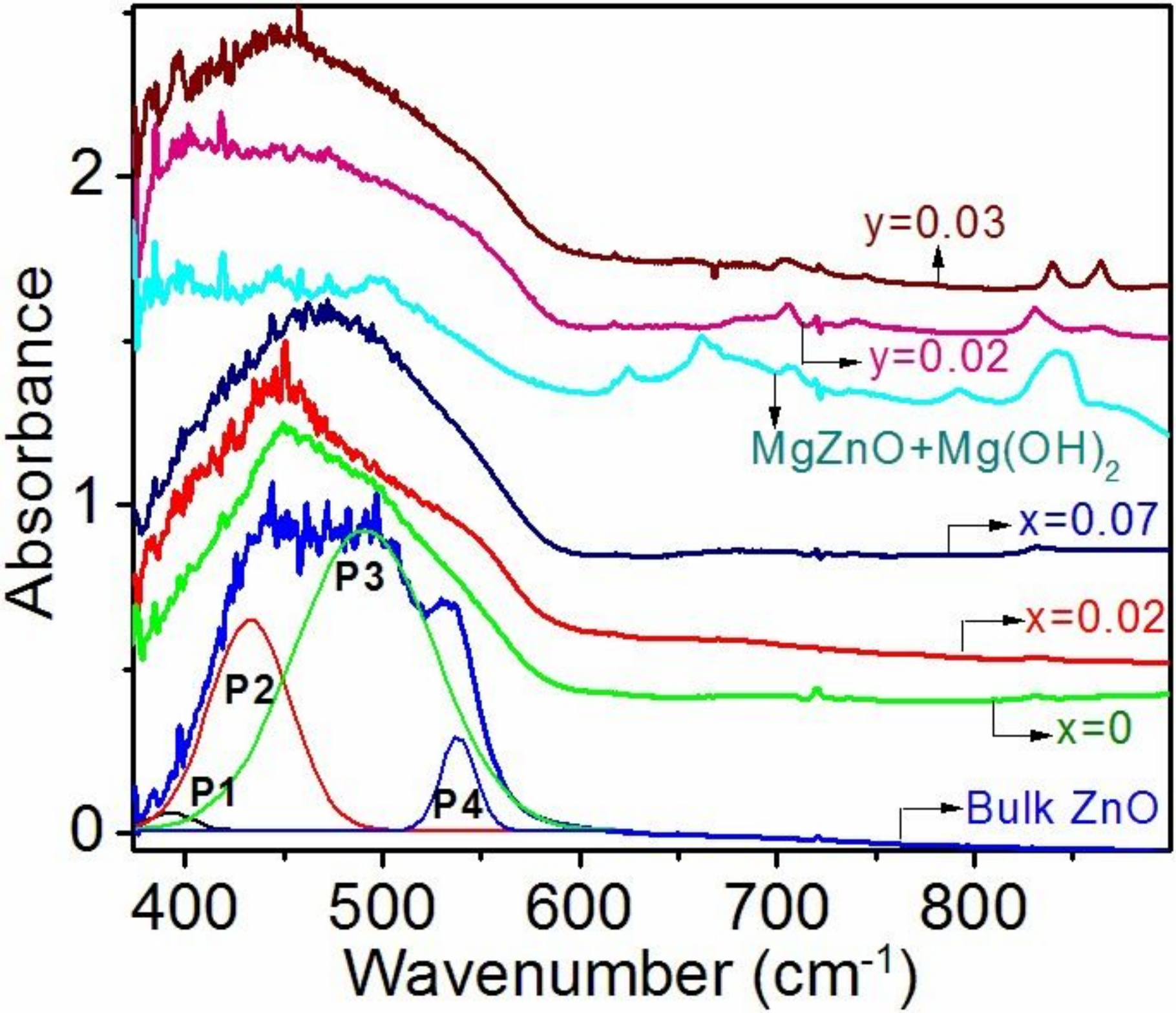

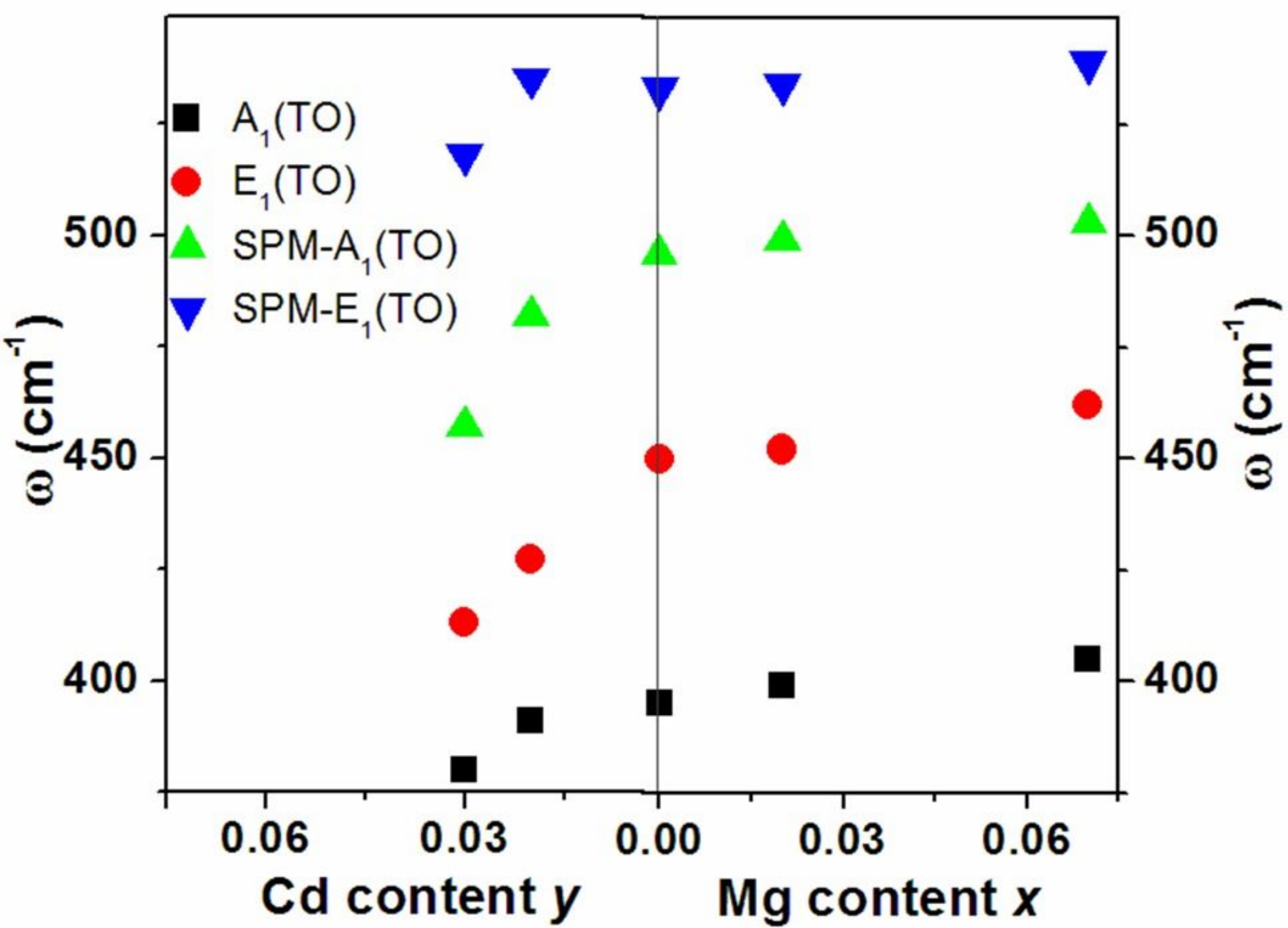